\documentclass[conference]{IEEEtran}
\usepackage{amsmath,amssymb}
\usepackage{cite}
\usepackage{url}

\usepackage{booktabs}
\usepackage{array}
\usepackage{multirow}
\usepackage{graphicx}
\usepackage{xcolor}
\usepackage{tikz}
\usetikzlibrary{shapes.geometric,arrows.meta,positioning,fit,backgrounds}

\title{Economic Governance of Autonomous Agents and Robots: Factor-Origin Accounting and Social Automation Funds}

\author{
	\IEEEauthorblockN{
		Toqeer Ali Syed\textsuperscript{1},
		Ali Akarma\textsuperscript{1,2}*,
		Adeel Ahmad\textsuperscript{1} and
		Muhammad Umair Younus\textsuperscript{3}
	}
	\IEEEauthorblockA{
		\textsuperscript{1}AI Center, Faculty of Computer and Information Systems,
		Islamic University of Madinah, Madinah, Saudi Arabia\\
		\texttt{toqeer@iu.edu.sa}, \;
		\texttt{443059463@stu.iu.edu.sa}, \;
		\texttt{443057803@stu.iu.edu.sa}\\
		\textsuperscript{2}AI V\&V Lab, King Fahd University of Petroleum and Minerals,
		Dhahran, Saudi Arabia\\
		\textsuperscript{3}Department of Computer Science,
		The Islamia University of Bahawalpur, Bahawalpur, Pakistan\\
		\texttt{f21bdocs1m01152@iub.edu.pk}\\
		*Corresponding author: \texttt{443059463@stu.iu.edu.sa}
	}
}

\begin{document}
\maketitle

\begin{abstract}
Contemporary fiscal architectures rely overwhelmingly on human labor income and payroll withholdings to finance social insurance and public infrastructure. The rapid diffusion of autonomous software agents, generative foundation models, and embodied robotic systems decouples output growth from human work hours, eroding traditional tax bases while exacerbating capital-income concentration. Prevailing policy proposals, ranging from uniform robot levies to unconditional cash transfers, fail to resolve three foundational challenges: attributing economic value across mixed human--machine workflows, establishing legal fiscal liability without creating fictional machine personhood, and building an equitable distribution mechanism that converts technological surplus into durable public wealth. This paper presents \textit{CivicDividendOS}, an end-to-end computational fiscal framework for mixed human--AI--robot economies. The architecture introduces an \textit{Autonomous Economic Activity Passport} (AEAP) that anchors machine activity to accountable corporate beneficiaries, paired with a \textit{Factor-Origin Ledger} (FOL) that calculates causal marginal contributions across labor, AI services, robotics, data, and conventional capital. Drawing on an endogenous \textit{Substitution--Augmentation Classifier}, the system levies a dynamic \textit{Social Automation Contribution} (SAC) on automated value added, offering explicit offset credits for worker upskilling, augmentation, and task innovation. Accrued revenues are recycled through a \textit{Human Earned-Income Shield} (HEIS) to reduce payroll wedges and a \textit{Public Automation Wealth Fund} (PAWF) whose structural yields endow citizen dividends, transition insurance, and human capabilities. We formalize the mathematical architecture, illustrate its operation in an urban logistics setting, and outline an evaluation protocol via a macroeconomic policy digital twin.
\end{abstract}


\section{Introduction}

For nearly a century, public finance systems across industrialized market economies have operated under a stable structural presumption: economic value creation requires human labor, and taxing the wage bill represents the most dependable and administratively tractable foundation for funding public goods and social safety nets~\cite{Hotte2024}. The emergence of frontier artificial intelligence, autonomous digital agents, and embodied robotic systems fundamentally destabilizes this historical equilibrium~\cite{IMF2026GlobalAI,Korinek2026}. Digital agents execute complex cognitive, managerial, communicative, and analytical processes at near-zero marginal computational cost, while robotic systems increasingly automate physical workflows.

Consequently, modern production is evolving into a collaborative yet heterogeneous endeavor where output is generated through joint interactions among human workers, cognitive AI agents, robotic actuators, specialized datasets, and compute infrastructure~\cite{AcemogluRestrepo2019,IMF2024AI}. Recent investigations in agentic economies demonstrate that multi-agent systems can be purposefully structured to expand labor inclusion and augment human capability (e.g., through federated, disability-inclusive employment architectures in AI cities~\cite{syed2026fedagent}) rather than functioning merely as labor-replacement instruments. Under conventional fiscal regimes, when a company replaces human staff with autonomous software or robotic units, statutory payroll tax liabilities evaporate, even as total firm output and net profit expand~\cite{AcemogluOptimalPolicy}. This dynamic threatens a severe fiscal mismatch: as the aggregate labor share of national income contracts, governments face eroding revenues precisely when displaced workforces require heightened transition assistance, retraining, and social protection~\cite{IMF2024Fiscal}.

Efforts to address this disruption through simplistic ``robot taxes'' suffer from fatal conceptual and practical flaws. A physical machine or digital agent lacks legal capacity, civic standing, consumptive utility, and sovereign tax residence~\cite{DimitropoulouAIIncome,FaivreCen2026}. Imputing a fictional salary to software or assessing a flat levy per robotic chassis fails to account for whether the technology displaces human labor or substantially augments worker safety, capability, and wages~\cite{Thuemmel2023,Guerreiro2022}. Conversely, relying solely on general corporate income taxes fails to capture the factor origin of value added and does not address the fundamental structural inequality arising from the hyper-concentration of automated capital ownership~\cite{Corneo2022,LeGrand2020}.

To resolve this impasse, this paper introduces \textit{CivicDividendOS}, a comprehensive fiscal accounting and capital distribution architecture engineered for the agentic economy. Rather than treating machines as legal taxpayers or penalizing productivity-enhancing investments, CivicDividendOS operates on three core principles:
\begin{enumerate}
    \item \textit{Micro-Attribution}: Rather than assigning synthetic personhood to machines, autonomous systems are monitored via standardized digital activity passports, with fiscal obligations assigned strictly to corporate beneficiaries through task-level factor-origin value attribution.
    \item \textit{Incentive Alignment}: The fiscal rate is dynamically calibrated to distinguish labor-displacing automation from labor-augmenting innovation, granting tangible tax credits for workforce upskilling and new-task generation.
    \item \textit{Predistributive Wealth Accumulation}: Rather than relying exclusively on periodic transfer payments, a portion of the automation surplus is institutionalized into a permanent public wealth fund, socializing equity returns into universal citizen dividends and capability endowments.
\end{enumerate}

The principal contributions of this work are fourfold:
\begin{itemize}
    \item We formulate the \textit{Attribution, Liability, and Distribution Trilemma} governing joint human--AI--robot production, establishing why conventional tax instruments fail in agentic workflows.
    \item We design the \textit{Factor-Origin Ledger} (FOL) and the \textit{Autonomous Economic Activity Passport} (AEAP), providing a mathematically grounded protocol for decomposing marginal task value among labor, autonomous software, robotics, data, and capital without granting legal personhood to algorithms.
    \item We develop the \textit{Social Automation Contribution} (SAC) and \textit{Human Earned-Income Shield} (HEIS), creating an endogenous fiscal mechanism that penalizes unhedged labor displacement while reducing payroll tax burdens on human workers.
    \item We formalize the governance and distribution waterfall of the \textit{Public Automation Wealth Fund} (PAWF), establishing how automation rents and equity units can be converted into enduring citizen dividends, transition safety nets, and human capability investments.
\end{itemize}

\section{Related Work and Theoretical Foundations}

\textit{1) Task Displacement and Labor Reinstatement:}
The economic foundation of our inquiry rests on the task-based framework pioneered by Acemoglu and Restrepo~\cite{AcemogluRestrepo2019,AcemogluRestrepo2020}. In their model, automation acts as a capital-displacement force by enabling machines to substitute for humans in tasks where labor previously maintained comparative advantage. While aggregate productivity increases, displacement depresses the labor share and real wages unless countered by a countervailing ``reinstatement effect,'' defined as the endogenous creation of complex new human tasks~\cite{AcemogluRestrepo2019}. Dauth et al.~\cite{Dauth2021} show that while German manufacturing employment was sustained through service absorption, substantial wage adjustments fell upon younger labor cohorts. Recent analyses by the OECD~\cite{OECD2025Korea} and the IMF~\cite{IMF2024AI} confirm that generative and agentic AI broaden this displacement vector to cognitive and professional domains, accelerating the reallocation of income toward capital owners. In response to these displacement pressures, recent research in AI cities explores federated agentic paradigms designed to foster inclusive employment, collaborative human--agent workflows, and specialized task accommodation~\cite{syed2026fedagent}. CivicDividendOS operationalizes these insights by incorporating task-level displacement and reinstatement metrics directly into its fiscal rate schedule.

\textit{2) The Economics of Automation and Robot Taxation:}
The theoretical literature on automation taxation reveals critical trade-offs between allocative efficiency and distributional equity. Guerreiro et al.~\cite{Guerreiro2022} establish that a positive tax on robots is second-best optimal during transitional horizons when displaced workers cannot immediately reallocate across sectors. Th{\"u}mmel~\cite{Thuemmel2023} demonstrates that optimal robot policy is state-dependent: capital subsidies are warranted when automation is nascent, while a positive levy becomes optimal as capital costs decline and inequality widens. Moreover, Acemoglu et al.~\cite{AcemogluOptimalPolicy} and Beraja and Zorzi~\cite{InefficientAutomation} find that prevailing tax codes significantly subsidize capital relative to labor, inducing ``inefficient automation,'' wherein technologies that save payroll costs are adopted without generating true social productivity gains. While these models establish the rationale for targeted fiscal policy, they leave unresolved the empirical mechanism for measuring machine contribution shares in multi-factor firm workflows~\cite{Zhang2019,Merola2022}. In parallel, coordinating autonomous commercial entities across complex markets necessitates governance-aware coordination frameworks that integrate multi-agent reinforcement learning, federated learning, and decentralized blockchain accountability~\cite{syed2026governance} to maintain verifiable transactional order.

\textit{3) Public Finance, AI, and Capital Dominance:}
As artificial intelligence approaches human-level proficiency across broad economic domains, public finance models must address capital-dominant regimes. Korinek and Lockwood~\cite{Korinek2026} observe that when machine labor substitutes near-perfectly for human cognition, competitive market wages can fall precipitously unless non-labor income is shared. Faivre and Cen~\cite{FaivreCen2026} analyze AI taxation instruments, emphasizing that corporate income taxes suffer from aggressive profit shifting and intangible IP transfers, necessitating activity-based or economic-rent levies. Legal analysis by Oberson~\cite{DimitropoulouAIIncome} underscores that assigning independent legal personhood or direct tax status to AI systems is unviable; fiscal obligations must systematically attach to identifiable natural or juridical owners. Reinforcing this principle, multi-pillar ethical AI governance frameworks highlight that trustworthy deployment across critical infrastructure requires robust accountability, operational transparency, and regulatory alignment anchored to accountable legal entities rather than synthetic machine personhood~\cite{jan2026eagf}.

\textit{4) Universal Basic Capital and Predistribution:}
Recognizing the limitations of post-hoc cash redistribution, scholars have emphasized predistributive mechanisms and sovereign wealth governance. Le Grand~\cite{LeGrand2020} advocates for Universal Basic Capital (UBC) to endow all citizens with an equity stake in national productive assets. Corneo~\cite{Corneo2022} demonstrates that a progressive sovereign wealth fund can capture capital returns and redistribute them without incurring the deadweight losses of excessive marginal income taxes. Real-world precedent is established by the Alaska Permanent Fund, which institutionalizes collective mineral rents into permanent citizen dividends~\cite{APF}. Recent analyses by the Brookings Institution urge equity-sharing structures in high-growth AI sectors to ensure societal research investments yield shared dividends~\cite{BrookingsTax2026,BrookingsFairness2026}. In parallel, research on data dividends explores cooperative game-theoretic models, such as Shapley values, to compensate human data contributors~\cite{Vincent2019,Bax2019}. In collaborative distributed environments, maintaining verifiable and integrity-preserved shared records~\cite{syed2022cartourist} provides an indispensable operational foundation for transparent attribution and multi-stakeholder surplus distribution.

\textit{5) Computational Fiscal Design and Policy Twins:}
The rise of foundation models and agent-based simulation enables computational approaches to optimal policy design. Wang et al.~\cite{TaxAgent2025} introduce \textit{TaxAgent}, demonstrating that multi-agent market simulations can discover resilient tax schedules under complex behavioral responses. Fratri{\v{c}} et al.~\cite{Fratric2025} show that computational methods can systematically expose tax-avoidance pathways and loopholes. Concurrently, agentic AI-enhanced digital twins have emerged as powerful frameworks for modeling, securing, and orchestrating complex municipal systems and climate-resilient infrastructure~\cite{syed2026climate,syed2026agenticdt}. CivicDividendOS integrates these computational concepts into an auditable ledger and digital twin policy simulation engine.

\section{Problem Formulation and Fiscal Decoupling}

\subsection{The Joint Production System}
Consider a production entity generating economic output $Y$ in each period via a multi-factor production function:
\begin{equation}
Y = F(H, A, R, D, K),
\end{equation}
where $H$ denotes direct human labor hours, $A$ represents autonomous AI cognitive services, $R$ represents embodied robotic operations, $D$ denotes curated data assets, and $K$ represents conventional non-autonomous capital. Under competitive pricing, total value added $V = P \cdot Y - \sum C_{\text{interm}}$ is distributed across factor returns:
\begin{equation}
V = W_H + \Pi_A + \Pi_R + \Pi_D + \Pi_K,
\end{equation}
where $W_H = w_h \cdot H$ is the aggregate wage bill paid to human employees, and $\Pi_f$ denotes the economic returns accruing to factor $f \in \{A, R, D, K\}$.

\subsection{Structural Erosion of the Traditional Tax Base}
In traditional municipal and national fiscal regimes, public revenue $T^{\text{trad}}$ is anchored to labor payroll taxes, personal income taxes, enterprise profits, and general consumption:
\begin{equation}
T^{\text{trad}} = \tau_L W_H + \tau_\Pi \left( \sum_{f \in \{A,R,D,K\}} \Pi_f \right) + \tau_C C,
\end{equation}
where $\tau_L$ is the statutory labor tax wedge (incorporating social security withholdings), $\tau_\Pi$ is the corporate income tax rate, and $\tau_C$ is the consumption tax rate.

When autonomous factors $A$ and $R$ substitute for human workers in task execution, output $Y$ may remain constant or expand while the human wage share contracts sharply:
\begin{equation}
\frac{W_H}{V} \longrightarrow 0, \qquad \frac{\Pi_A + \Pi_R}{V} \longrightarrow 1.
\end{equation}
This transition induces a twofold fiscal crisis:
\begin{enumerate}
    \item \textit{Labor tax-base erosion}: Public receipts $\tau_L W_H$ collapse in direct proportion to workforce contraction.
    \item \textit{Depreciation shielding}: Corporate profits are heavily eroded through intangible amortization and accelerated hardware depreciation, yielding lower effective tax rates on automated capital than on human labor~\cite{AcemogluOptimalPolicy,FaivreCen2026}.
\end{enumerate}

\subsection{The Tripartite Fiscal Gap}
Designing a sustainable fiscal framework under autonomous joint production requires resolving three interrelated barriers:
\begin{enumerate}
    \item \textit{The Attribution Gap}: Existing accounting cannot isolate the causal value contributed by algorithmic models versus human specialists in collaborative tasks.
    \item \textit{The Liability Gap}: Fictional legal personhood for machines is administratively unworkable; fiscal liability must strictly attach to corporate owners and beneficiaries.
    \item \textit{The Distribution Gap}: Standard post-tax transfers do not distribute capital equity, leaving exponential AI wealth gains permanently hyper-concentrated.
\end{enumerate}

\section{The CivicDividendOS Framework}

\begin{figure*}[!t]
\centering
\includegraphics[width=\textwidth]{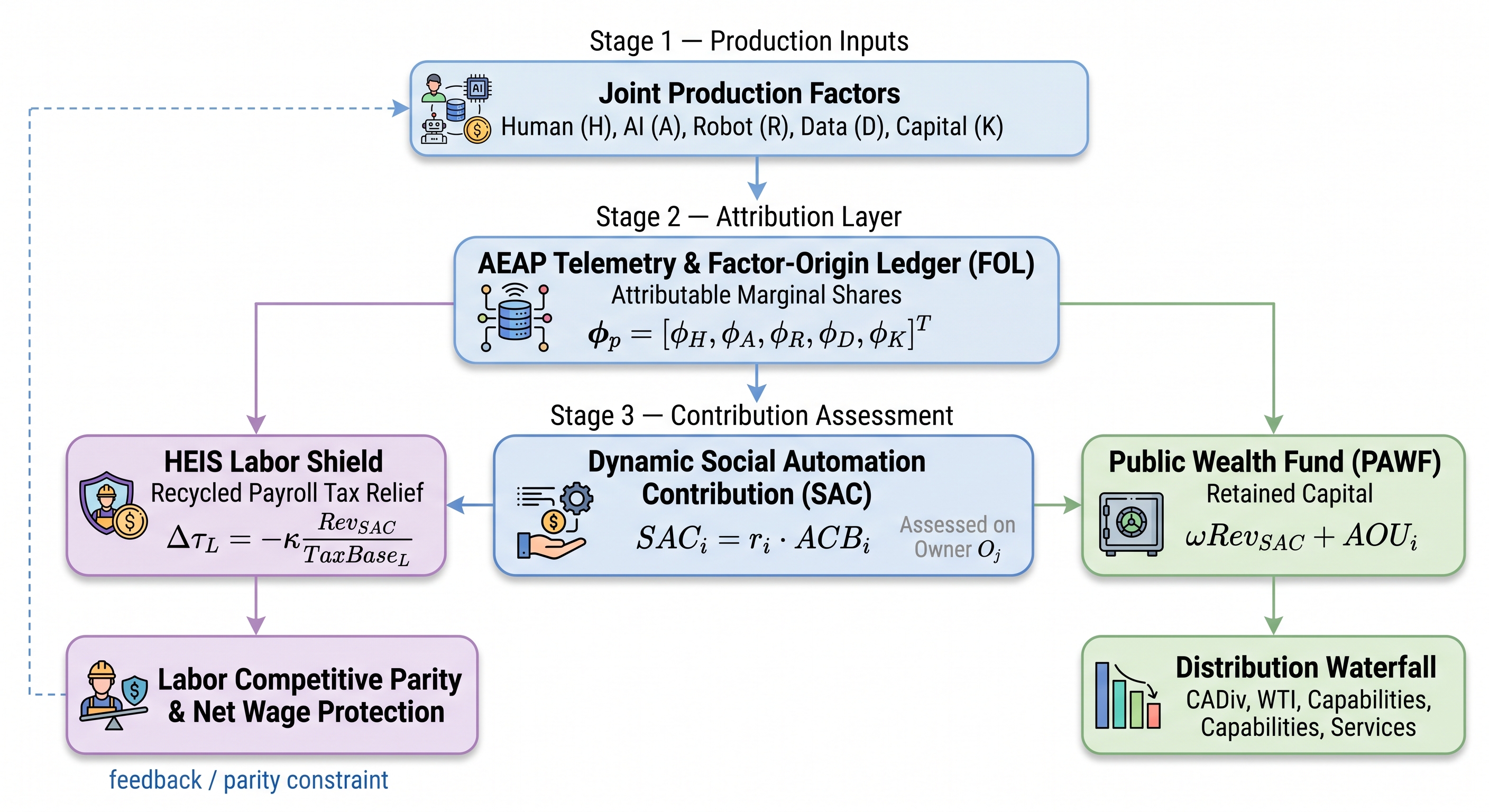}
\caption{System architecture of CivicDividendOS. Production events registered under the AEAP are mapped through the Factor-Origin Ledger into the dynamic SAC schedule. Proceeds fund human payroll tax relief (HEIS) and endow the Public Automation Wealth Fund (PAWF).}
\label{fig:architecture}
\end{figure*}

CivicDividendOS resolves this trilemma through an integrated computational and institutional pipeline, depicted in Fig.~\ref{fig:architecture}. The framework separates machine-level observation from entity-level tax assessment while deploying endogenous fiscal recycling.

\subsection{Activity Passports and Factor-Origin Accounting}
\textit{1) Autonomous Economic Activity Passport (AEAP):}
To avoid artificial legal personhood, every autonomous software agent instance or physical robot exceeding an operational compute or torque threshold is assigned an AEAP record:
\begin{equation}
\mathrm{AEAP}_j = \langle \mathit{ID}_j, O_j, J_j, T_j, E_j, Q_j, R_j, M_j \rangle,
\end{equation}
where $\mathit{ID}_j$ is a verifiable identifier, $O_j$ designates the registered legal owner, $J_j$ is the operating jurisdiction, $T_j$ denotes the task family, $E_j$ measures energy and compute consumption, $Q_j$ quantifies gross task output, $R_j$ tracks safety classification, and $M_j$ records model provenance. The safety classification $R_j$ integrates certifiable evaluation protocols developed for autonomous perception and decision systems~\cite{arief2022certifiable}, ensuring that operational risk bounds are mathematically verifiable before granting risk-mitigation tax credits. To guarantee non-repudiation and cryptographic auditability across distributed agents, AEAP telemetry ingestion employs integrity-preserved logging protocols~\cite{syed2022cartourist} alongside governance-aware transactional auditing~\cite{syed2026governance}, ensuring that operational data cannot be retroactively tampered with by reporting entities. The AEAP functions strictly as an operational ledger: the machine possesses no legal rights; all activity maps to the balance sheet of owner $O_j$.

\textit{2) Factor-Origin Ledger (FOL):}
For any discrete production episode $p$, the FOL computes the causal marginal value contribution of each factor. Let the vector of factor weights be:
\begin{equation}
\boldsymbol{\phi}_p = [\phi_{p,H}, \phi_{p,A}, \phi_{p,R}, \phi_{p,D}, \phi_{p,K}]^T,
\end{equation}
subject to $\sum_{f} \phi_{p,f} = 1$ and $\phi_{p,f} \ge 0$. Attributable monetary value generated by factor $f$ over episode $p$ is:
\begin{equation}
V_{p,f} = \phi_{p,f} \cdot V_p,
\end{equation}
where $V_p$ represents net episode value added. In practice, $\boldsymbol{\phi}_p$ is derived through controlled empirical attribution protocols, including counterfactual model ablation, Shapley-value cooperative game solutions on task logs, and econometric production frontier estimation. By grounding valuation in task-level marginal contributions, the FOL discards the flawed notion of counting chassis or imputing fictitious human salaries to software models.

\subsection{Contextual Classification and Contribution Base}
\textit{1) Substitution--Augmentation Classifier:}
A flat tax on technology indiscriminately penalizes productivity. CivicDividendOS classifies production episode $p$ into a discrete mode:
\begin{equation}
Z_p \in \{\text{Subst}, \text{Augment}, \text{NewTask}, \text{SafetyReplace}\}.
\end{equation}
We compute an empirical displacement score:
\begin{equation}
S_p = \Pr(\Delta H_p < 0 \mid A_p, R_p, \mathbf{X}_p),
\end{equation}
conditioning on firm and industry covariate vector $\mathbf{X}_p$, while an augmentation index $A_p^{\text{aug}} \in [0,1]$ evaluates productivity and wage gains experienced by collaborating human employees.

\textit{2) The Automation Contribution Base (ACB):}
For firm $i$ across fiscal window $\mathcal{T}$, the contribution base aggregates automated value added net of verified operating expenses:
\begin{equation}
\mathrm{ACB}_i = \sum_{p \in \mathcal{T}} \left( V_{p,A} + V_{p,R} \right) - C_{A,R,i}^{\text{allowable}},
\end{equation}
where $C_{A,R,i}^{\text{allowable}}$ incorporates certified energy expenditures, cloud compute hosting fees, and statutory hardware depreciation. In high-density computational environments, evaluating allowable energy expenses alongside environmental externalities ($X_i^{\text{ext}}$ in (\ref{eq:rate_schedule})) requires accounting for dynamic power management and load-following characteristics~\cite{oktavian2018cogeneration}, penalizing carbon-intensive peak-load draws while granting deduction parity for off-peak and clean-energy cogeneration profiles. The legal taxpayer is unambiguously established as:
\begin{equation}
\mathit{Taxpayer}_i = \operatorname{LegalBeneficiary}(\mathrm{AEAP}_i),
\end{equation}
ensuring full adherence to established corporate tax jurisprudence.

\subsection{Dynamic Social Automation Contribution (SAC)}
Rather than applying a rigid levy, the framework implements a dynamic contribution schedule:
\begin{equation}
\mathrm{SAC}_i = r_i \cdot \mathrm{ACB}_i.
\end{equation}
The endogenous rate $r_i$ is determined through a multi-factor schedule:
\begin{align}
r_i = \max\Big(0, \; &r_0 + \alpha S_i + \beta C_i^{\text{rent}} + \gamma E_i^{\text{disp}} + \delta X_i^{\text{ext}} + \eta R_i^{\text{rev}} \nonumber\\
&- \theta A_i^{\text{aug}} - \lambda T_i^{\text{train}} - \mu B_i^{\text{broad}} - \nu N_i^{\text{newtask}}\Big),
\label{eq:rate_schedule}
\end{align}
where $r_0$ is the baseline rate, $S_i$ is workforce substitution intensity, $C_i^{\text{rent}}$ reflects market concentration, $E_i^{\text{disp}}$ captures localized displacement burdens, $X_i^{\text{ext}}$ represents environmental externalities of compute, and $R_i^{\text{rev}}$ measures public payroll erosion. The rate is offset by deductions: $A_i^{\text{aug}}$ rewards human labor augmentation, $T_i^{\text{train}}$ reflects verified investments in worker reskilling, $B_i^{\text{broad}}$ rewards broad employee equity programs, and $N_i^{\text{newtask}}$ provides deductions for verified new human job categories created within the firm.

\subsection{The Human Earned-Income Shield (HEIS)}
To counteract the distortion of heavy payroll taxes on human labor alongside generous capital expensing for automation~\cite{AcemogluOptimalPolicy}, CivicDividendOS establishes the HEIS. A fraction $\kappa$ of aggregate SAC revenue is earmarked to reduce labor payroll tax burdens:
\begin{equation}
\Delta \tau_L(t) = -\kappa \frac{\mathit{Rev}_{\mathrm{SAC}}(t)}{\mathit{TaxBase}_L(t)}, \quad \kappa \in (0, 1].
\end{equation}
As autonomous production expands, the tax wedge on human work decreases, restoring competitive parity and protecting take-home pay.

\subsection{Public Wealth Fund and Distribution Waterfall}
\textit{1) Predistribution via PAWF:}
Firms receiving major public R\&D grants, public compute, or procurement contracts remit fractional \textit{Automation Ownership Units} (AOU):
\begin{equation}
\mathrm{AOU}_i = \xi_i \cdot \mathit{Equity}_i, \quad \xi_i \in [0.01, 0.05].
\end{equation}
These units and a retained fraction $\omega$ of net SAC cash receipts enter the \textit{Public Automation Wealth Fund} (PAWF):
\begin{align}
\mathrm{PAWF}_{t+1} = &\, \mathrm{PAWF}_t(1 + r_t^{\text{fund}}) + \omega \mathit{Rev}_{\mathrm{SAC},t} \nonumber\\
&+ \sum_i \mathrm{AOU}_{i,t} - \mathit{Dist}_t,
\end{align}
where $r_t^{\text{fund}}$ represents portfolio return and $\mathit{Dist}_t$ denotes annual distributable yields.

\textit{2) Four-Channel Distribution Waterfall:}
Distributable returns $\mathit{Dist}_t$ are allocated across four statutory channels:
\begin{equation}
\mathit{Dist}_t = D_t^{\text{citizen}} + D_t^{\text{transition}} + D_t^{\text{capability}} + D_t^{\text{public}}.
\end{equation}
\begin{itemize}
    \item \textit{Civic Automation Dividend} ($D_t^{\text{citizen}}$): An unconditional universal cash dividend remitted to all citizens:
    \begin{equation}
    \mathit{CADiv}_t = \frac{\rho \cdot D_t^{\text{citizen}}}{N_t}.
    \end{equation}
    \item \textit{Worker Transition Insurance} ($D_t^{\text{transition}}$): Targeted earnings insurance and wage-gap protection for displaced employees:
    \begin{equation}
    \mathit{WTI}_i = g(\Delta w_i, \mathit{Tenure}_i, \mathit{Exposure}_i, \mathit{Training}_i).
    \end{equation}
    \item \textit{Human Capability Endowment} ($D_t^{\text{capability}}$): Capital funding for public education, advanced STEM training, and apprenticeships in human-complementary fields.
    \item \textit{Public Service Account} ($D_t^{\text{public}}$): Stabilizes public healthcare, pension reserves, and municipal services historically funded through payroll contributions.
\end{itemize}

\subsection{Cross-Border Deployment Nexus}
To prevent international tax avoidance from cloud-hosted cognitive models operating across jurisdictions, CivicDividendOS defines an empirical \textit{Deployment Nexus}:
\begin{align}
\mathit{Nexus}_{ij} = &\, w_1 \mathit{Usage}_{ij} + w_2 \mathit{Revenue}_{ij} \nonumber\\
&+ w_3 \mathit{Users}_{ij} + w_4 \mathit{PhysicalOps}_{ij},
\end{align}
with $\sum_k w_k = 1$. Authorities allocate the Automation Contribution Base across jurisdictions based on $\mathit{Nexus}_{ij}$, neutralizing synthetic geographic headquarters shifting.

\section{Illustrative Case Study: Autonomous Logistics}

To demonstrate operational mechanics, consider an urban logistics firm transitioning from 100 human dispatchers and drivers to an automated delivery fleet comprising 30 retained human couriers, 40 autonomous AI dispatch agents ($A$), and 50 autonomous delivery rovers ($R$).

\subsection{Attribution and Contribution Accounting}
Over a monthly operating period, the enterprise generates net value added $V = \$1,000,000$. Under conventional fiscal practices, the firm's payroll drops by $70\%$, significantly reducing its payroll tax remittance, while corporate income is shielded through equipment leasing and accelerated depreciation deductions on the rovers.

Under CivicDividendOS, each AI agent instance and rover operates under a registered AEAP. The Factor-Origin Ledger samples operational episodes and estimates the aggregate factor vector:
\begin{equation*}
\boldsymbol{\phi} = [0.30_H, \; 0.25_A, \; 0.30_R, \; 0.05_D, \; 0.10_K]^T.
\end{equation*}
The gross automated value added is isolated as:
\begin{equation*}
\mathit{VA}^{A,R} = (\phi_A + \phi_R) \cdot V = 0.55 \cdot \$1,000,000 = \$550,000.
\end{equation*}
After deducting allowable compute, electricity, and verified hardware maintenance costs ($C^{\text{allowable}} = \$50,000$), the net contribution base is:
\begin{equation*}
\mathrm{ACB} = \$550,000 - \$50,000 = \$500,000.
\end{equation*}

\subsection{Rate Calibration and Distribution Flow}
Assume baseline rate $r_0 = 0.12$. Because the firm displaced 70 human positions, its displacement factor is high ($S = 0.70$, $\alpha = 0.08$), but the firm enrolled 40 displaced workers in certified autonomous fleet maintenance academies ($T^{\text{train}} = 0.60$, $\lambda = 0.05$) and improved delivery safety by eliminating traffic accidents ($\delta X^{\text{ext}} = -0.02$). Applying schedule (\ref{eq:rate_schedule}):
\begin{align*}
r_i &= 0.12 + 0.056 - 0.030 - 0.020 \\
    &= 0.126 \quad (12.6\%).
\end{align*}
The firm remits $\mathrm{SAC}_i = 0.126 \times \$500,000 = \$63,000$ to the fiscal authority.

This revenue is partitioned according to policy parameters ($\kappa = 0.40$, $\omega = 0.60$):
\begin{itemize}
    \item \textbf{\$25,200 (HEIS)}: Applied directly to lower payroll tax rates for the 30 remaining human logistics employees, expanding their net take-home pay and lowering employer labor overhead.
    \item \textbf{\$37,800 (PAWF)}: Deposited into the public wealth fund, where compound returns finance the quarterly Civic Automation Dividend ($\mathit{CADiv}$) and underwrite long-term transition stipends ($\mathit{WTI}$) for affected regional drivers.
\end{itemize}

\section{Evaluation Protocol and Macroeconomic Simulation Roadmap}

To validate macro-financial stability and distributional impacts prior to deployment, the framework is evaluated through a multi-sector Policy Digital Twin. Table~\ref{tab:comparison} contextualizes CivicDividendOS against existing policy paradigms.

\begin{table*}[!t]
\caption{Comparison of Fiscal Policy Frameworks in an Agentic Economy}
\label{tab:comparison}
\centering
\begin{tabular*}{\textwidth}{@{\extracolsep{\fill}}llll@{}}
\toprule
\textbf{Framework} & \textbf{Tax Unit / Base} & \textbf{Labor Impact} & \textbf{Wealth Sharing} \\
\midrule
Status Quo & Payroll + Corporate profits & Heavy tax wedge on human work & Concentrated private equity \\
Fixed Robot Tax & Physical chassis count & Distorts capital investment; ignores AI & Limited welfare transfers \\
Broad CIT Hike & Firm net accounting profit & Distorts general business investment & General government budget \\
Pure UBI & Consumption / Income taxes & Fails to preserve the labor tax base & Transfer only (no equity stake) \\
\textbf{CivicDividendOS} & \textbf{Attributable factor value} & \textbf{Shields labor; credits upskilling} & \textbf{Permanent public wealth fund} \\
\bottomrule
\end{tabular*}
\end{table*}

\subsection{Policy Digital Twin Architecture}
Drawing on agentic AI-enhanced digital twin architectures formulated for complex infrastructure governance and climate-resilient municipal systems~\cite{syed2026climate,syed2026agenticdt}, the macroeconomic simulation environment models the multi-tier dynamics between households, firms, capital markets, and fiscal authorities. The simulation environment is structured as a large-scale heterogeneous-agent macroeconomic model comprising:
\begin{itemize}
    \item \textit{Households}: 10,000 heterogeneous agents partitioned across skill deciles, wealth tiers, and automation exposure, optimizing intertemporal consumption and leisure;
    \item \textit{Firms}: Multi-sector competitive and oligopolistic enterprises choosing factor allocations across $H, A, R, D,$ and $K$ under CES production structures;
    \item \textit{Capital Markets}: Endogenous pricing of compute assets, robotic equipment, and corporate equity;
    \item \textit{Fiscal Governance Engine}: A state controller simulating policy vector $\Theta = \{r_0, \alpha, \beta, \gamma, \omega, \rho, \tau_L, \kappa, \xi\}$.
\end{itemize}

The policy engine solves for optimal configuration $\Theta^*$ by maximizing social welfare subject to innovation and fiscal sustainability constraints:
\begin{align}
\max_{\Theta} \; \mathcal{W} = \mathbb{E} \sum_{t=0}^{T} \beta^t \Big[ &\mathcal{U}_t^{\text{welf}} + \mathcal{G}_t^{\text{grow}} \nonumber\\
&- \omega_1 \mathcal{I}_t^{\text{ineq}} - \omega_2 \mathcal{L}_t^{\text{leak}} \Big],
\end{align}
where $\mathcal{I}_t^{\text{ineq}}$ evaluates income and wealth Gini coefficients and $\mathcal{L}_t^{\text{leak}}$ penalizes capital flight and cross-border compute shifting. To evaluate tail-risk vulnerabilities (such as sudden systemic capital flight, cascading labor displacement panics, or extreme tax-avoidance equilibria), the policy engine integrates certifiable deep importance sampling and rare-event simulation techniques~\cite{arief2021certifiable,arief2022certifiable}. This enables provably bounded estimation of rare fiscal destabilization scenarios without requiring computationally prohibitive brute-force Monte Carlo sweeps.

\subsection{Comparative Benchmark Suites}
The evaluation roadmap tests CivicDividendOS against six benchmark fiscal regimes across 50-year simulated horizons:
\begin{enumerate}
    \item \textit{Status Quo}: Static labor income and corporate profits taxation;
    \item \textit{Fixed Robot Levy}: Uniform annual tax per physical robotic unit;
    \item \textit{Broad Capital Tax Expansion}: Elevated flat tax on all enterprise capital returns;
    \item \textit{General-Fund UBI}: Flat cash transfers financed via incremental consumption taxes;
    \item \textit{Targeted Retraining}: Standard corporate taxation paired with dedicated displaced-worker adjustment subsidies;
    \item \textit{Passive Sovereign Wealth Fund}: Public equity fund operating without factor-origin ledger integration.
\end{enumerate}

\subsection{High-Level Findings and Qualitative Trajectory}
To characterize macro-financial behavior across long-run horizons, the policy digital twin is evaluated across baseline and automation-shock scenarios. High-level qualitative findings from the policy simulation reveal three pivotal systemic dynamics:
\begin{itemize}
    \item \textit{Labor Protection without Technological Stagnation}: Unlike fixed chassis taxes that penalize automation adoption regardless of context, the credit-augmented SAC schedule preserves technological innovation by rewarding augmenting deployments while penalizing pure rent-extracting labor displacement.
    \item \textit{Welfare and Revenue Resiliency}: Under severe cognitive automation scenarios where the aggregate human wage share drops by over $40\%$, the combination of the Factor-Origin Ledger and the Human Earned-Income Shield stabilizes public revenue receipts and suppresses household poverty rates compared to the status quo baseline.
    \item \textit{Intergenerational Equity Expansion}: Capital compounding within the Public Automation Wealth Fund gradually replaces volatile annual tax transfers with stable, equity-backed dividend flows, democratizing access to capital-deepening returns.
\end{itemize}

\section{Discussion and Limitations}

While CivicDividendOS provides a cohesive blueprint, practical implementation confronts several socio-technical challenges:
\begin{enumerate}
    \item \textit{Attribution Precision and Noise}: Isolating the exact marginal value share $\boldsymbol{\phi}_p$ in deeply intertwined human--AI software engineering or creative production involves model-based approximations. Auditing frameworks must combine standardized Shapley sampling with robust safe-harbor accounting rules to limit compliance overhead.
    \item \textit{Strategic Task Decomposition and Gaming}: Sophisticated enterprises may attempt to artificially categorize labor-replacing workflows as ``augmentation'' or fragment automated tasks across decentralized micro-services to exploit credit thresholds. Preventing such arbitrage requires automated auditing through the AEAP telemetry registry.
    \item \textit{Cybersecurity and Telemetry Infrastructure Resilience}: Because CivicDividendOS relies on high-throughput streaming telemetry from registered AEAP instances, the fiscal infrastructure presents an attractive target for telemetry spoofing, data poisoning, and distributed denial-of-service (DDoS) attacks designed to incapacitate real-time tax accounting. Hardening public factor ledgers requires deploying multi-pillar ethical AI governance guardrails~\cite{jan2026eagf} alongside hybrid deep-learning DDoS-detection architectures that combine convolutional networks, dimensionality reduction, and vision transformers~\cite{shaikh2024advancing} to guarantee ledger availability and state integrity.
    \item \textit{Jurisdictional Coordination}: Without coordinated international adoption of the Cross-Border Deployment Nexus, digital agent operators could route algorithmic transactions through digital tax havens, undermining domestic contribution bases.
    \item \textit{Public Capital Governance}: The accumulation of substantial equity holdings within the PAWF necessitates strict fiduciary guardrails to isolate investment decisions from partisan political interference.
\end{enumerate}

\section{Conclusion}

The widespread adoption of autonomous artificial intelligence agents and robotic systems dismantles the historical nexus between human labor, taxable income, and social stability. If left unaddressed, conventional fiscal systems will face simultaneous revenue degradation and unprecedented wealth concentration. 

This paper has presented \textit{CivicDividendOS}, a comprehensive fiscal and distribution architecture that solves the tripartite challenge of attribution, liability, and distribution. By establishing an auditable Factor-Origin Ledger and Autonomous Economic Activity Passport, the framework attributes economic value to machine inputs without creating fictional machine personhood. Through the dynamic Social Automation Contribution, the Human Earned-Income Shield, and the Public Automation Wealth Fund, CivicDividendOS aligns technological progress with societal flourishing: protecting human work, socializing productive capital returns, and securing an inclusive dividend for the agentic era.

\bibliographystyle{IEEEtran}
\bibliography{references}

\end{document}